\documentclass[pre,aps,preprint,showpacs,preprintnumbers,amsmath,amssymb,secnumroman,eqsecnum]{revtex4}

\usepackage{graphicx}
\usepackage{dcolumn}
\usepackage{bm}

\newcommand{\beq}{\begin{equation}}
\newcommand{\eeq}{\end{equation}}
\newcommand{\beqa}{\begin{eqnarray}}
\newcommand{\eeqa}{\end{eqnarray}}
\newcommand{\vc}[1]{\mbox{\boldmath $#1$}}

\newcommand{\vol}[1]{{\bf #1}}

\newcommand{\du}[1]{{\bf\sf #1}}

\begin{document}


\title{Optimization of flagellar swimming by a model sperm}

\author{B. U. Felderhof}

 \email{ufelder@physik.rwth-aachen.de}
\affiliation{Institut f\"ur Theorie der Statistischen Physik \\ RWTH Aachen University\\
Templergraben 55\\52056 Aachen\\ Germany\\
}%

\date{\today}

\begin{abstract}
The swimming of a bead-spring chain in a viscous incompressible fluid as a model of a sperm is studied in the framework of low Reynolds number hydrodynamics. The optimal mode in the class of planar flagellar strokes of small amplitude is determined on the basis of a generalized eigenvalue problem involving two matrices which can be evaluated from the mobility matrix of the set of spheres constituting the chain. For an elastic chain with a cargo constraint for its spherical head, the actuating forces yielding a nearly optimal stroke can be determined. These can be used in a Stokesian dynamics simulation of large amplitude swimming.
\end{abstract}

\maketitle
\section{\label{I}Introduction}

The sperm of most animals swim by means of a planar flagellar stroke \cite{1},\cite{1A}. A sperm consists of a tail and a head, the latter in many cases having a larger diameter than the tail. In order to estimate the swimming speed for any given power in the framework of low Reynolds number hydrodynamics \cite{2} it is important to take full account of hydrodynamic interactions. In the early work of Gray and Hancock \cite{3} and of Lighthill \cite{4} the calculations were simplified by the assumption that the swimming velocity of a headless string could be used to find the corresponding thrust needed to move the head, balancing it with the drag on the head as evaluated in infinite fluid. Such a calculation does not take proper account of hydrodynamic interference effects between tail and head \cite{4A}. Higdon performed an improved calculation for both planar \cite{5} and helical \cite{6} flagellar swimming which took approximate account of such effects, but he assumed a plane wave stroke. We find that the optimal stroke is rather different.

In the following we study planar flagellar swimming for a bead-spring chain model with actuating forces \cite{7}. In order to understand the hydrodynamics of swimming it is essential to mimic the flow as accurately as possible. We are not concerned with explaining the motion of the tail in terms of its actual structure \cite{1},\cite{1A}, or with understanding the molecular basis of its mechanism \cite{8}.

By concentrating on the hydrodynamic aspects of the problem we are able to find the optimum mode of small amplitude swimming. In the earlier work the stroke was assumed to have plane wave character, although Gray and Hancock \cite{3} had already remarked that observations of the actual stroke show significant deviations from a plane wave. End effects are important. The numerical calculations assuming a plane wave showed maximum efficiency for a wavelength corresponding to approximately one wave over the length of the organism \cite{5}.

The bilinear theory of swimming, as developed for a general assembly of spheres \cite{9},\cite{9A} allows the calculation of the optimum stroke of small amplitude. For any stroke one can find the corresponding actuating forces in the bead-spring model which are necessary to perform it. For a chain with a head we impose a cargo constraint, implying that the actuating force on the head vanishes, so that the force on the head is caused only by direct elastic interactions with the beads. We find that for long chains the constraint does not prevent the stroke from having near optimal swimming efficiency. For larger amplitudes we amplify the actuating forces and calculate the swimming speed and power from the limit cycle of the motion calculated from Stokesian dynamics \cite{10}.

A continuum model with bending and stretching forces and with simplified hydrodynamics and some form of resistive force theory has been studied in much of the earlier work \cite{1},\cite{11}-\cite{11G}. The discrete bead-spring model can be conveniently used in computer simulations. Gauger and Stark \cite{12} have studied the swimming of a chain with Rotne-Prager hydrodynamic interactions actuated by an applied magnetic field and by induced magnetic dipole interactions. Flexible swimmers without a head with Oseen monopole hydrodynamic interactions have been studied by Llopis et al. \cite{13}.

In Sec. II of this paper we discuss the bead-spring chain model in some detail, and explain the strategy of the subsequent calculations. In Sec. III we study chains without a head. In Sec. IV we discuss the complications resulting from the presence of a head. In Sec. V we consider the effects of a cargo constraint. The article ends with a Discussion.

\section{\label{II}Swimming bead-spring chain}

In the following we consider a microorganism consisting of a tail and a spherical head or cell body, swimming in a viscous incompressible fluid of shear viscosity $\eta$. The organism is propelled through the fluid by action of the tail. We are primarily interested in a proper description of the hydrodynamics, and choose a simplified description of the tail. The latter will be modeled as a bead-spring chain consisting of $N-1$ identical spheres of radius $a$ linked by harmonic interactions for deviations from equally spaced equilibrium positions. The spherical head is linked to the last bead by harmonic interaction permitting elastic vibrations about its own equilibrium position. The equilibrium positions are located sequentially on the $x$ axis of a Cartesian coordinate system. We consider two limiting instances, one where the head is much larger than the beads, and one where the head is identical with a bead or equivalently, where the chain is headless. In Fig. 1 we show as an example a chain consisting of three beads and a head of radius $b=2.5a$ with the beads moving in the transverse direction.

We assume that the motion of the chain of $N$ spheres is caused by actuating forces $\vc{E}_1(t),...,\vc{E}_N(t)$ satisfying the constraint
\begin{equation}
\label{2.1}\sum^N_{j=1}\vc{E}_j(t)=\vc{0}.
\end{equation}
The additional assumption that the head is passive is expressed by the cargo constraint $\vc{E}_N(t)=\vc{0}$. An $N$-chain with the cargo constraint is denoted as an $NC$-chain. The forces on the individual spheres are a sum of actuating and elastic interaction forces. For an $NC$-chain the force on the $N$th sphere is due to elastic interaction only. For longitudinal excitation we consider actuating forces in the $x$ direction and study the motion of the $x$ coordinates of the sphere centers. For transverse planar excitation we consider actuating forces in the $y$ direction and restrict attention to the motion of centers in the $xy$ plane.

In low Reynolds number hydrodynamics the flow velocity $\vc{v}$ and pressure $p$ of the fluid are assumed to satisfy the Stokes equations
\begin{equation}
\label{2.2}\eta\nabla^2\vc{v}-\nabla p=0,\qquad\nabla\cdot\vc{v}=0.
\end{equation}
The fluid moves due to a no-slip boundary condition on the surface of each sphere. The solution of the hydrodynamic problem for known forces $(\vc{F}_1,...,\vc{F}_N)$ exerted by the spheres on the fluid can be expressed by a mobility matrix $\vc{\mu}$ relating the sphere translational velocities $(\vc{U}_1,...,\vc{U}_N)$ linearly to the forces. In abbreviated form the relation reads
\begin{equation}
\label{2.3}\du{U}=\vc{\mu}\cdot\du{F}.
\end{equation}
No torques are exerted, so that it is not necessary to consider the rotational velocities of the spheres. The mobility matrix depends on the positions of the sphere centers $\du{R}=(\vc{R}_1,..,\vc{R}_N$), but only via relative distance vectors $\vc{r}_{ij}=\vc{R}_i-\vc{R}_j$ due to translational invariance.

The forces $\du{F}$ exerted on the fluid are related to the actuating forces $\du{E}$ by
\begin{equation}
\label{2.4}\du{F}=\du{E}+\du{H}\cdot(\du{R}-\du{S}_0),
\end{equation}
where the matrix $\du{H}$ is constant, symmetric, and with the property that the elastic force vector vanishes for a uniform shift of the equilibrium positions $\du{S}_0$ without change of relative positions. The positions $\du{R}$ change in time according to the equations of Stokesian dynamics \cite{14}
\begin{equation}
\label{2.5}\frac{d\du{R}}{dt}=\du{U}(\du{R}).
\end{equation}
The actuating forces $\du{E}$ are assumed to vary harmonically in time with frequency $\omega=2\pi/T$, where $T$ is the period. The solution of Eq. (2.5) tends to a limit cycle. The mean swimming velocity $\overline{\vc{U}}_{sw}$ is identical to the mean velocity of any of the spheres in the final cyclic motion. The average is over a period of the motion. We are also interested in the mean power, corresponding to the rate of dissipation in the fluid. The instantaneous rate of dissipation is $\mathcal{D}=\du{F}\cdot\du{U}$.

In swimming the total hydrodynamic force vanishes at all times. In our case this can be expressed as
\begin{equation}
\label{2.6}\vc{F}_N(t)=-\sum^{N-1}_{j=1}\vc{F}_j(t).
\end{equation}
The condition is satisfied on account of Eq. (2.1) and Newton's third law for the interaction forces. In the resistive force theory of Gray and Hancock \cite{3} and the slender body theory of Lighthill \cite{4},\cite{15} the equation is interpreted as thrust$=-$drag. The right hand side is calculated for a headless chain, and the swimming velocity is calculated from Stokes' law $\vc{U}_N\approx\vc{F}_N/(6\pi\eta b)$, where $b$ is the radius of the big sphere. It is clear that these calculations are only approximate, and that hydrodynamic interactions are not properly accounted for.

In the calculations of Gray and Hancock \cite{3} and Lighthill \cite{4} the stroke of the headless string is assumed to be given. In the resistive force theory the corresponding hydrodynamic forces exerted by the string are calculated from assumed longitudinal and transverse friction coefficients. Lighthill \cite{4} has provided an improved estimate of the friction coefficients. In the slender body theory one evaluates the hydrodynamic forces exerted by the headless string from integral equations with approximate kernels. Subsequently one evaluates the mean swimming velocity and mean power, assuming the equivalent of Eq. (2.6) and using Stokes' law. The assumed motion is of plane wave type, and the mean swimming velocity and mean power are evaluated as a function of wavenumber. The efficiency, i.e. the ratio of mean speed and mean power, can be optimized as a function of wavenumber. Typically, for a finite string the number of waves for optimum swimming for this type of stroke is of order unity. Higdon \cite{5},\cite{6} employed improved integral equations with hydrodynamic interactions taking account of the head, but he assumed a plane wave stroke.

Our strategy is different. We use the bilinear theory of swimming to optimize the swimming efficiency of an $N$-chain with a head. From the corresponding relative displacement vector and assuming harmonic interactions we can calculate the first $N-1$ actuating forces $\vc{E}_j(t)$. Assuming the cargo constraint $\vc{E}_N=\vc{0}$ and modifying $\vc{E}_{N-1}(t)$ such that Eq. (2.1) is satisfied, we can calculate the swimming mode and mean swimming velocity and mean power of the $NC$-chain. The corresponding efficiency is slightly less than the optimum efficiency of the $N$-chain without constraint. The virtue of this approach is that it results in a good estimate of the best set of actuating forces satisfying the cargo constraint. The same set of actuating forces, multiplied by an amplitude factor, can be used in a Stokesian dynamics calculation. The mean swimming velocity and mean power can be evaluated from the numerically determined limit cycle. It may reasonably be expected that this procedure provides a good stroke for large amplitude swimming. The numerical calculations for finite chains with a head show that both for longitudinal squirming and for planar flagellar swimming the stroke is substantially different from a plane wave, in that there is a significantly larger amplitude at the tail end than near the head.

In our calculations we use approximate expressions for the mobility matrix. For a headless chain we use Oseen monopole interactions for simplicity. The calculation could be improved by use of Rotne-Prager pair interactions \cite{16},\cite{17}. This would provide better estimates of swimming speed and power, but would make no difference in principle. For a chain with head we use the recently developed approximate mobility matrix for small spheres and a big sphere \cite{18}, based on Oseen's exact Green function for a fluid in the presence of a sphere with no-slip boundary condition \cite{19}.

\section{\label{III}Actuating forces}

In this section we discuss the construction of an appropriate set of actuating forces in more detail. We look for solutions of eq. (2.5) corresponding to swimming motion, of the form
\begin{equation}
\label{3.1}\vc{R}_j(t)=\vc{S}_{j0}+\int^t_0\vc{U}(t')\;dt'+\vc{\delta}_j(t),\qquad j=1,...,N,
\end{equation}
where the first two terms describe the collective motion of the equilibrium configuration $\du{S}_0=(\vc{S}_{10},...,\vc{S}_{N0})$ with instantaneous swimming velocity $\vc{U}(t)$ caused by the displacements $\{\vc{\delta}_j(t)\}$. We require that the latter are periodic with period $T$, and exclude uniform displacements, so that the $3N$-dimensional vector $\du{d}(t)=\{ \vc{\delta}_1(t),...,\vc{\delta}_N(t)\}$ satisfies
\begin{equation}
\label{3.2}\du{d}(t)\cdot\du{u}_\alpha=0,\qquad(\alpha=x,y,z),
\end{equation}
where the symbol $\du{u}_x$ denotes a $3N$-dimensional vector with $1$ on the $x$ positions, $0$ on the $y,z$ positions, and cyclic.

The optimization of strokes of small amplitude is performed in relative space. Therefore we transform to center and relative coordinates. The center of the assembly is given by
\begin{equation}
\label{3.3}\vc{R}=\frac{1}{N}\sum_{j=1}^N\vc{R}_j=\frac{1}{N}\;\vc{e}_\alpha\du{u}_\alpha\cdot\du{R}
\end{equation}
with Cartesian unit vectors $\vc{e}_\alpha$ and summation over repeated greek indices implied. We define relative coordinates $\{\vc{r}_j\}$ as
  \begin{eqnarray}
\label{3.4}\vc{r}_1&=&\vc{R}_2-\vc{R}_1,\qquad\vc{r}_2=\vc{R}_3-\vc{R}_2,\qquad ...,\nonumber\\
\vc{r}_{N-1}&=&\vc{R}_N-\vc{R}_{N-1}, \qquad j=1,...,N-1.
\end{eqnarray}
and the corresponding $(3N-3)$-vector $\du{r}=(\vc{r}_1,...,\vc{r}_{N-1})$. The $3N$-vector $(\vc{R},\du{r})$ is related to the vector $\du{R}$ by a transformation matrix $\du{T}$ according to
\begin{equation}
\label{3.5}(\vc{R},\du{r})=\du{T}\cdot\du{R}
\end{equation}
with explicit form given by eqs. (3.3) and (3.4). The instantaneous swimming velocity equals the rate of change of the center, $\vc{U}(t)=d\vc{R}/dt$.

Displacements $\vc{\xi}$ in relative space are defined by
\begin{equation}
\label{3.6}(\vc{0},\vc{\xi})=\du{T}\cdot\du{d}.
\end{equation}
In the bilinear theory, corresponding to small $\du{d}$, the orbit in relative space is given by $\vc{r}(t)=\vc{r}_0+\vc{\xi}_0(t)$, with $\vc{r}_0$ given by the relative position vector in the equilibrium configuration, and
\begin{equation}
\label{3.7}\vc{\xi}_0(t)=\varepsilon a\;\mathrm{Re}\;\vc{\xi}^c_0\exp(-i\omega t),
\end{equation}
with amplitude factor $\varepsilon$, and a selected $(3N-3)$-dimensional vector $\vc{\xi}^c_0$. The superscript $c$ indicates that the vector components are dimensionless complex numbers.  The corresponding displacement vector in configuration space is given by
\begin{equation}
\label{3.8}\du{d}_0(t)=\du{T}^{-1}\cdot\left(\begin{array}{c}\vc{0}\\\vc{\xi}_0(t)\end{array}\right).
\end{equation}

By series expansion in powers of $\varepsilon$ we obtain a corresponding expansion of the instantaneous swimming velocity
\begin{equation}
\label{3.9}\vc{U}=\vc{U}^{(1)}+\vc{U}^{(2)}+\vc{U}^{(3)}+...,
\end{equation}
with first order term
\begin{equation}
\label{3.10}U^{(1)}_{0\alpha}=-M^0_{\alpha\beta}\du{u}_\beta\cdot\vc{\zeta}^0\cdot\dot{\du{d}_0},
\end{equation}
with mobility tensor $M^0_{\alpha\beta}$ and friction matrix $\vc{\zeta}^0$ calculated for the configuration $\du{S}_0$. By periodicity of $\du{d}_0(t)$ the time average of the first order swimming velocity vanishes, $\overline{\vc{U}_0^{(1)}}=\vc{0}$.

The first order forces $\du{F}^{(1)}_0(t)$ corresponding to the displacement vector $\du{d}_0(t)$ are given by
\begin{equation}
\label{3.11}\du{F}^{(1)}_0=\vc{\zeta}^0\cdot(U^{(1)}_{0\beta}\du{u}_\beta+\dot{\du{d}}_0).
\end{equation}
The corresponding actuating forces $\du{E}_0(t)$ are found from eq. (2.4) as
 \begin{equation}
\label{3.12}\du{E}_0(t)=\du{F}^{(1)}_0(t)-\du{H}\cdot\du{d}_0(t).
\end{equation}
These have the property $\du{u}_\alpha\cdot\du{E}_0(t)=0$, so that the sum of actuating forces vanishes.

 We recall that in the bilinear theory of swimming the mean rate of dissipation and the mean swimming velocity can be expressed  \cite{9},\cite{9A} as expectation values of a power matrix $\du{A}$ and a speed matrix $\du{B}^\alpha$ with respect to an amplitude vector $\vc{\xi}^c$. Thus to second order in the amplitude of the stroke the mean swimming velocity in the $x$ direction and the mean rate of dissipation are given by
 \begin{equation}
\label{3.13}\overline{U}_{sw2}=\frac{1}{2}\omega
a(\vc{\xi}^c|\du{B}^x|\vc{\xi}^c),\qquad\overline{\mathcal{D}}_2=
\frac{1}{2}\eta\omega^2a^3(\vc{\xi}^c|\du{A}|\vc{\xi}^c).
 \end{equation}
The elements of the matrices $\du{A}$ and $\du{B}^x$ are dimensionless and can be evaluated from the $N$-sphere Stokes mobility matrix \cite{9},\cite{9A}. The scalar product is defined as
  \begin{equation}
\label{3.14}(\vc{\xi}^c|\vc{\eta}^c)=\sum^{N-1}_{j=1}\vc{\xi}_j^{c*}\cdot\vc{\eta}^c_j.
 \end{equation}

The swimming efficiency is defined in general as
 \begin{equation}
\label{3.15}E_T=\eta\omega a^2\frac{|\overline{U}_{sw}|}{\overline{\mathcal{D}}}.
 \end{equation}
To second order in the amplitude this may be evaluated from Eq. (3.13). Optimization of the second order efficiency gives rise to the generalized eigenvalue problem
 \begin{equation}
\label{3.16}\du{B}^x\vc{\xi}^c_\lambda=\lambda\du{A}\vc{\xi}^c_\lambda.
 \end{equation}

By axial symmetry of the equilibrium configuration the matrices $\du{A}$ and $\du{B}^x$ decompose as
 \begin{equation}
\label{3.17}\du{A}=\du{A}^x\oplus\du{A}^y\oplus\du{A}^z,\qquad\du{B}^x=\du{B}^{xx}\oplus\du{B}^{xy}\oplus\du{B}^{xz},
\end{equation}
where $\du{A}^x,\;\du{A}^y,\;\du{A}^z$ and $\du{B}^{xx},\;\du{B}^{xy},\;\du{B}^{xz}$ are $(N-1)\times (N-1)$-dimensional and $\;\du{A}^y=\;\du{A}^z$, $\;\du{B}^{xy}=\;\du{B}^{xz}$.
In the case of longitudinal motion along the $x$ axis it is sufficient to consider the $(N-1)\times (N-1)$-dimensional hermitian matrices $\;\du{A}^x$ and $\;\du{B}^{xx}$.  We denote the $(N-1)$-dimensional eigenvector with maximum eigenvalue for longitudinal motion by $\vc{\xi}^x_0$. This can be completed to a $(3N-3)$-dimensional vector to be used in Eq. (3.7) by adding zeros for the $y$ and $z$ components. It is clear by symmetry that the corresponding actuating forces $\du{E}^x_0(t)$ have vanishing $y$ and $z$ components.

For transverse first order motion in the $y$ direction perpendicular to the axis and swimming in the $x$ direction it is sufficient to consider the $(N-1)\times (N-1)$-dimensional hermitian matrices $\;\du{A}^y$ and $\;\du{B}^{xy}$. The swimming speed and power for planar flagellar motion can be calculated from the matrices $\du{A}^y,\;\du{B}^{xy}$. Optimization of the swimming speed for fixed power leads to the generalized eigenvalue problem
 \begin{equation}
\label{3.18}\du{B}^{xy}\vc{\xi}^c=\lambda^y\du{A}^y\vc{\xi}^c,
 \end{equation}
for the case of planar flagellar motion. We denote the $(N-1)$-dimensional eigenvector with maximum eigenvalue for this eigenvalue problem by $\vc{\xi}^y_0$.  This can be completed to a $(3N-3)$-dimensional vector to be used in Eq. (3.7) by adding zeros for the $x$ and $z$ components. It is clear by symmetry that the corresponding actuating forces $\du{E}^y_0(t)$ have vanishing $x$ and $z$ components.

\section{\label{IV}Headless chains}

In this section we consider headless chains of identical beads with Oseen monopole interactions. In a previous paper \cite{10} we considered longitudinal motions of the beads along the chain axis. Here we first derive analytical results for short chains swimming in the direction of the axis caused by transverse flagellar motion of the beads. In the construction of the matrices $\;\du{A}^y$ and $\;\du{B}^{xy}$ one can use two different procedures \cite{9},\cite{9A}. In the second procedure one needs the gradient of the $3N\times 3N$-dimensional friction matrix $\vc{\zeta}(\du{R})$ in configuration space. This is calculated conveniently from the gradient of the mobility matrix $\vc{\mu}(\du{R})$ by use of $\vc{\zeta}\cdot\vc{\mu}=\du{I}$, which yields
  \begin{equation}
\label{4.1}\vc{\nabla}\vc{\zeta}=-\vc{\zeta}\cdot\vc{\nabla}\vc{\mu}\cdot\vc{\zeta},
\end{equation}
where the contractions refer to the components of $\vc{\zeta}$ and $\vc{\mu}$. The relation is needed at $\du{S}_0$, and at this point in configuration space the friction matrix $\vc{\zeta}^0=\vc{\zeta}(\du{S}_0)$ can be found in explicit form on account of symmetry.

The simplest example of planar flagellar swimming is a linear chain of three equal-sized spheres of radius $a$ with equal distance $d $ between centers 1,2 and 2,3, and mobility matrix calculated in Oseen monopole approximation. The matrices $\du{A}$ and $\du{B}^x$ are six-dimensional, and decompose as in Eq. (3.17). The $2\times2$ matrices $\du{A}^x$ and $\du{B}^{xx}$ were calculated earlier from the purely longitudinal motion \cite{9}. The matrix $\du{B}^{xy}$ takes the form
  \begin{equation}
\label{4.2}\du{B}^{xy}=\left(\begin{array}{cc}0&iaY^y_{12}
\\-iaY^y_{12}&0
\end{array}\right),
\end{equation}
with element
  \begin{equation}
\label{4.3}Y^y_{12}=\frac{-a}{3d}\frac{112d^2-306da+189a^2}{(8d-3a)(8d-7a)(4d-7a)}.
\end{equation}
The matrix $\du{A}^y$ takes the form
 \begin{equation}
\label{4.4}\du{A}^y=\frac{16\pi d}{(8d-3a)(8d-7a)}\left(\begin{array}{cc}16d-12a&8d-9a
\\8d-9a&16d-12a
\end{array}\right).
\end{equation}
These expressions are to be compared with those for longitudinal motion \cite{9}. The matrix $\du{B}^{xy}$ shows a singularity at $d=7a/4$, which is less than the minimum distance $2a$. From Eq. (4.4) one finds the eigenvalues
\begin{equation}
\label{4.5}\lambda^y_\pm=\mp\frac{a}{16\sqrt{3}\pi d}\sqrt{(8d-3a)(8d-7a)}Y^y_{12},
\end{equation}
 as well as the corresponding eigenvectors $\vc{\xi}^y_\pm=(1,\xi^y_\pm)$ with
 \begin{equation}
\label{4.6}\xi^y_+=\frac{1}{16d-12a}\bigg[-8d+9a+i\sqrt{3(8d-3a)(8d-7a)}\bigg],\qquad\xi^y_-=\xi_+^{y*},
\end{equation}
normalized to $(\vc{\xi}^y_+|\vc{\xi}^y_+)=2$. The maximum efficiency, corresponding to $\lambda^y_+$, tends to zero monotonically as the ratio $d/a$ tends to infinity. The eigenvalue $\lambda^y_+$ for transverse motion is less than the corresponding $\lambda^x_+$ for longitudinal motion for all values of the ratio $d/a$. The ratio $\lambda^x_+/\lambda^y_+$ tends to 2 in the limit $d/a\rightarrow\infty$. Hence for this 3-chain the maximum efficiency for longitudinal motion is always larger than for transverse motion.

For values of $N$ larger than 4 the algebraic solution becomes too cumbersome, and one must resort to numerical methods. In particular, an inverse matrix must be calculated and in the construction of the matrix $\du{B}^{xy}$ the derivative with respect to positions must be taken numerically. We confirm numerically that the decomposition Eq. (3.17) holds. Hence the eigenvalue problem for the longitudinal and transverse modes  can be considered separately.

In Fig. 2 we plot the efficiency $E_{Tmax}$, as given by the maximum eigenvalue $\lambda_{max}$ for $N=3,...,16$ for a chain of $N$ spheres of radius $a$ with distance $d=5a$ between successive sphere centers in the rest configuration and with monopole Oseen interactions, for both the longitudinal and transverse mode. It is seen that for these values of $N$ the longitudinal mode is always the most efficient. The efficiency decreases with increasing $N$.

In order to find the wave of sphere displacements corresponding to the optimum eigenvector we must use the transformation Eq. (3.8). In Fig. 3 we plot the $x$ components of the displacement vector $\du{d}_0(t)$ corresponding to the eigenvector $\vc{\xi}^{x}_0$ with maximum eigenvalue for the longitudinal mode for $N=12$, normalized such that $\xi^{x}_{01}=1$. We show the displacements $\delta_{0jx}(t)$ at times $t=0,\;t=\frac{1}{8}T,\;t=\frac{1}{2}T,\;t=\frac{3}{8}T,\;t=\frac{1}{2}T$, with the discrete amplitudes connected by a continuous curve found by interpolation. In Fig. 4 we show the corresponding curves for the $y$ components of the displacement vector $\du{d}_0(t)$ corresponding to the eigenvector $\vc{\xi}^{y}_0$ with maximum eigenvalue for the transverse mode for $N=12$, normalized such that $\xi^{y}_{01}=1$.

\section{\label{V}Chains with a head}

In this section we consider chains of identical beads of radius $a$ with a head of radius $b$. We use hydrodynamic interactions derived in an approximation for $b>>a$ ($SB$- hydrodynamic interactions \cite{18}, $S$ stands for small. $B$ stands for big). In previous papers \cite{9},\cite{10} we considered longitudinal motions of the beads along the chain axis. Here we first derive analytical results for short chains swimming in the direction of the axis caused by transverse flagellar motion of the beads. Next we shall study numerical results for longer chains.

We consider first a 3-chain consisting of two spheres of radius $a$ and a third one of radius $b$ with $b>>a$. We assume that in the rest configuration the relative distances between centers on the $x$ axis are $(d,b+d)$ and that the spheres move only in the $xy$ plane. With $SB$-hydrodynamic interactions the analytic expressions for the matrices $\du{A}$ and $\du{B}$ become quite complicated. The matrices decompose as in Eq. (3.17). We present analytic results in the limit $a<<d<<b$. In this limit the matrices $\du{A}^x$ and $\du{B}^{xx}$ take the asymptotic form derived earlier \cite{10}. The matrices $\du{A}^y$ and $\du{B}^{xy}$ take the asymptotic form
\begin{equation}
\label{5.1}\du{A}^y_{as}=6\pi\left(\begin{array}{cc}1&1\\
1&2
\end{array}\right),\qquad\du{B}^{xy}_{as}=\frac{8a^3}{3b^3}\left(\begin{array}{cc}0&i\\
-i&0
\end{array}\right),\qquad (N=3).
\end{equation}
The matrix $\du{A}^y_{as}$ is identical to $\du{A}^x_{as}$, and in the matrix $\du{B}^{xy}_{as}$ the prefactor equals $8/3$, compared to $-15/4$ for the longitudinal mode. Hence the longitudinal mode is more efficient than the transverse one.

For $N=4$ with $SB$-hydrodynamic interactions we again find the decomposition Eq. (3.17). In the limit $a<<d<<b$ the matrix $\du{A}^y$ takes the asymptotic form
\begin{equation}
\label{5.2}\du{A}^y_{as}=6\pi\left(\begin{array}{ccc}1&1&1\\
1&2&2\\1&2&3
\end{array}\right),\qquad (N=4),
\end{equation}
and the matrix $\du{B}^{xy}$ becomes
 \begin{equation}
\label{5.3}\du{B}^{xy}_{as}=\frac{ia^3}{b^3}\left(\begin{array}{ccc}0&\frac{351}{50}&\frac{53703}{6400}\\
-\frac{351}{50}&0&\frac{3101}{768}\\-\frac{53703}{6400}&-\frac{3101}{768}&0
\end{array}\right),\qquad (N=4).
\end{equation}
The matrix $\du{A}^y_{as}$ is identical to $\du{A}^x_{as}$. From the generalized eigenvalue problem one shows that the maximum eigenvalue for the transverse mode is again smaller than for the longitudinal one.

In Fig. 5 we plot the maximum efficiency $E_{Tmax}$ for $N=3,...,16$ for a chain of $N-1$ spheres of radius $a$ with distance $d$ between successive sphere centers in the rest configuration, with a big sphere with label $N$ of radius $b$ with center at distance $b+d$ from the center of the $N-1$-th sphere for both the longitudinal and transverse mode for $d=5a,\;b=10a$. We see that for these chain lengths the longitudinal mode is the most efficient. For both the longitudinal and the transverse mode the efficiency is maximal at $N=11$.

In Fig. 6 we plot the $x$ components of the displacement vector $\du{d}_0(t)$ corresponding to the eigenvector $\vc{\xi}^{x}_0$ with maximum eigenvalue for the longitudinal mode for $N=11$, normalized such that $\xi^{x}_{01}=1$. In Fig. 7 we show the corresponding curves for the $y$ components of the displacement vector $\du{d}_0(t)$ corresponding to the eigenvector $\vc{\xi}^{y}_0$ with maximum eigenvalue for the transverse mode for $N=11$, normalized such that $\xi^{y}_{01}=1$.

The behavior in Figs. 6 and 7 shows that the optimal mode for a 11-chain with a head is approximately of the form $\xi^c_j=\exp(ik j-\gamma j)$. For a mode of this type we find in the longitudinal case for a chain of 11 spheres with $d=5a,\;b=10a$ efficiency $E_T=150\times 10^{-6}$ for $k=1.193,\;\gamma=0.193$, compared with the maximum value $E_{Tmax}=168\times 10^{-6}$, shown in Fig. 5. The presence of the head makes the absolute values $|\xi^x_{0j}|$ asymmetric about the midpoint. In the transverse case we find efficiency $E_T=77\times 10^{-6}$ for $k=1.197,\;\gamma=0.203$, compared with the maximum value $E_{Tmax}=86\times 10^{-6}$, shown in Fig. 5. In both cases we  have optimized the values of $k$ and $\gamma$.

\section{\label{VI}Elastic chains with cargo constraint}

The calculations performed so far have been purely kinematic. The bilinear theory of swimming allows determination of the optimum stroke of a specified type, but is not concerned with the question how the stroke can be achieved. In the following we consider the model discussed in Sec. II with actuating forces and harmonic elastic interactions.
For squirming and planar flagellar swimming the harmonic first order motion is assumed to be caused by actuating forces $\vc{E}_1(t),...,\vc{E}_N(t)$ with vanishing $z$ components and satisfying the constraint (2.1). The additional assumption that the head is passive is expressed by the cargo constraint $\vc{E}_N(t)=\vc{0}$. The forces on the individual spheres are a sum of actuating and elastic interaction forces. For an $NC$-chain the force on the $N$th sphere is due to elastic interaction only. For squirming swimming we consider actuating forces in the $x$ direction and study the motion of the $x$ coordinates of the sphere centers. For transverse planar excitation we look at actuating forces in the $y$ direction and restrict our attention to the motion of centers in the $xy$ plane.

We consider actuating forces varying harmonically in time and put
 \begin{equation}
\label{6.1}\vc{E}_j(t)=\mathrm{Re}\vc{E}_{j\omega}^ce^{-i\omega t},\qquad j=1,...,N.
\end{equation}
For the longitudinal mode we choose the vector $\vc{\xi}_0$ in Eq. (3.7) to be the eigenvector with maximum eigenvalue $\vc{\xi}^x_0$  and for the transverse mode we choose the eigenvector $\vc{\xi}^y_0$. The corresponding actuating forces are given by Eq. (3.12). We assume the elastic interaction to be isotropic. For $N=4$ we use a matrix $\du{H}$ of the form
\begin{equation}
\label{6.2}\du{H}=k\left(\begin{array}{cccc}-\vc{1}&\vc{1}&\vc{0}&\vc{0}\\\vc{1}&-2\;\vc{1}&\vc{1}&\vc{0}
\\\vc{0}&\vc{1}&-2\;\vc{1}&\vc{0}\\\vc{0}&\vc{0}&\vc{1}&-\vc{1}
\end{array}\right),\qquad (N=4),
\end{equation}
where $\vc{1}$ is the $3\times 3$ unit matrix and $\vc{0}$ is a $3\times 3$ matrix consisting of zeros. The corresponding form for $N=3$ and $N>4$ is obvious. The stiffness of the swimmer is characterized by the dimensionless number $\sigma=k/(\pi\eta a\omega)$.

If in addition we impose the condition that the big sphere is a passive cargo by requiring $\vc{E}_N=\vc{0}$, then there are only $N-2$ independent actuating forces, for which we can take, for example, $\vc{E}_1,...,\vc{E}_{N-2}$. We modify the actuating force on the last bead $\vc{E}_{N-1}$ in such a way that the total actuating force vanishes. We denote the actuating forces constructed in this manner as $\du{E}^C$. The big sphere does exert a hydrodynamic force on the fluid, but only due to the elastic interaction with its neighboring sphere labeled $N-1$.

We present results for the case $d=5a,\;b=10a$ with $SB$-hydrodynamic interactions. In Fig. 8 we compare the maximum efficiency of a $3C$-chain with flagellar motion with that for longitudinal squirming motion. For $\sigma=0$ the 3-chain with cargo constraint cannot swim, and the efficiency vanishes. In both the longitudinal and the transverse case the efficiency is maximal at an intermediate value of $\sigma$. For all values of $\sigma$ the efficiency for transverse motion is less than that for longitudinal motion. We found earlier \cite{10} that in the longitudinal case the efficiency is maximal at $\sigma_{0x}=9.054$ with efficiency $E^C_{T0x}=0.159E_{Tasx}$, where $E_{Tasx}=5a^3/(8\pi b^3)$. In the transverse case the efficiency is maximal at $\sigma_{0y}=8.086$ with efficiency $E^C_{T0y}=0.135E_{Tasy}$, where $E_{Tasy}=4a^3/(9\pi b^3)$ is the efficiency in the asymptotic limit $a<<d<<b$, as found from Eq. (5.1).

For longer chains we can determine the optimum actuating forces for an $NC$-chain with given value of the stiffness $\sigma$. The eigenvectors with maximum eigenvalue $\vc{\xi}^x_0$ and $\vc{\xi}^y_0$ can be calculated from the matrices $\du{A}$ and $\du{B}^x$ as in Sec. IV. The corresponding actuating forces $\vc{E}_1,...,\vc{E}_{N-1}$ for the unconstrained chain are calculated from Eqs. (3.10)-(3.12). Guided by the results for short chains we choose the value $\sigma=10$ in this calculation. Subsequently the actuating force $\vc{E}_{N-1}$ is modified such that Eq. (2.1) and the cargo constraint $\vc{E}_N=\vc{0}$ are satisfied. The corresponding displacement vector $\du{d}_\omega$ can be calculated from the linearized equations of Stokesian dynamics,
 \begin{equation}
\label{6.3}U^{(1)}_{\alpha\omega}\du{u}_\alpha-i\omega \du{d}_\omega=\vc{\mu}^0\cdot(\du{E}^C_\omega+\du{H}\cdot\du{d}_\omega),
\end{equation}
with use of Eq. (3.10). The corresponding vector $\vc{\xi}^C$ can be calculated from Eq. (3.6). For a long chain this does not differ much from the optimal vector $\vc{\xi}_0$, and correspondingly the efficiency is not much less than that of the unconstrained $N$-chain. For a 11C-chain with $d=5a,\;b=10a$ and $\sigma=10$ the efficiency for squirming motion is $E^C_T=1672\times10^{-7}$, compared with the optimum $E_{Tmax}=1679\times10^{-7}$ for the unconstrained 11-chain. For a 11C-chain with $d=5a,\;b=10a$ the efficiency for planar flagellar motion is $E^C_T=848\times10^{-7}$, compared with the optimum $E_{Tmax}=852\times10^{-7}$ for the unconstrained 11-chain. For such a long chain the cargo constraint has little effect on the efficiency. In Fig. 9 we present a plot of $(b^2/a^2)E_{Tmax}$ for the maximum efficiency of an unconstrained 11-chain as a function of the ratio $b/a$. This shows that at large $b/a$ the maximum speed at fixed power decays approximately as $a^2/b^2$.

Finally we show the motion of the first bead and the head for planar flagellar excitation. We consider again $d=5a,\;b=10a,\;\sigma=10$ and actuating forces calculated as indicated above for normalization $\xi^y_{01}=1$. The motion of the chain is calculated by numerical solution of the equations of Stokesian dynamics with the initial state at $t=0$ chosen to be the rest configuration. The motion tends to a limit cycle after a few periods. In Fig. 10 we show the $x$-component of the displacement from the equilibrium position of the first bead and the head of a $4C$-chain as a function of time for the first ten periods. In Fig. 11 we show the corresponding displacements in the $y$ direction. In Figs. 12 and 13 we show the same quantities for a $11C$-chain for the first twenty periods. From Figs. 10 and 12 the swimming motion is evident.

It is of interest to study the dependence on the amplitude of the forces. In Fig. 14 we show the reduced mean swimming velocity $\overline{U}_{sw}/\overline{U}_{sw2}$ of a $4C$-chain and a $11C$-chain as functions of $\log_{10}(\varepsilon)$ for the same actuating forces multiplied by $\varepsilon$, where $\overline{U}_{sw2}=\frac{1}{2}\varepsilon^2\omega a(\vc{\xi}^{Cy}|\du{B}^{xy}|\vc{\xi}^{Cy})$. This suggests that the mean swimming velocity varies quadratically over a wide range of $\varepsilon$. The same is true for the power. For the longitudinal motion one finds similar plots. The plots show that up to about $\varepsilon=1$ swimming velocity and power are found to be in agreement with the bilinear theory. Therefore the bilinear theory can be used with confidence to calculate the optimal actuating forces for a chain with cargo.

\section{\label{VII}Discussion}

The bead-spring chain model of a sperm allows for detailed calculations of the hydrodynamics of swimming motion. In particular, the theory of cyclic motion in the space of relative positions of sphere centers \cite{9}, and the corresponding construction of an eigenvalue problem permitting optimization of small amplitude motion, provide insights into the nature of the optimal stroke within the class of planar flagellar strokes. Presumably sperm in nature employ the optimal stroke. Interestingly, the calculations show that swimming by longitudinal squirming is more efficient than planar flagellar swimming. The bilinear theory provides a good estimate of the optimal stroke.

In the calculations an approximate mobility matrix with simplified hydrodynamic interactions has been used, but it can be improved, for example by the use of a Rotne-Prager pair interaction between beads. The mobility matrix of a chain of spheres can in principle be calculated accurately by use of the multipole expansion method \cite{20}-\cite{22}.

 A bead-spring model has advantages in computer simulation. The present theory provides the type of motion for study in Stokesian dynamics simulations. For example, it would be of interest to calculate the mean swimming velocity  for variations of the optimal stroke calculated for simplified hydrodynamic interactions. It would also be of interest to study the hydrodynamic interaction between two sperm swimming with the optimal planar flagellar stroke.

\newpage

\section*{Figure captions}

\subsection*{Fig. 1}
Schematic shape of a transverse planar swimmer consisting of three beads and a head.

\subsection*{Fig. 2}
Plot of the maximum efficiency $E_{Tmax}$ of a headless $N$-chain of identical beads of radius $a$ for rest distance $d=5a$ for longitudinal squirming (large dots) and transverse planar flagellar swimming (small dots) for $N=3,...16$.

\subsection*{Fig. 3}
Plot of the displacements $\delta_{jx}$ in the $x$ direction for a 12-chain of identical spheres with $d=5a$ for longitudinal squirming in the optimal mode, at times $t=0,\;t=T/8,\;t=T/4,\;t=3T/8,\;t=T/2$, labeled $1,...,5$. The points have been joined by interpolation for clarity.

\subsection*{Fig. 4}
Plot of the displacements $\delta_{jy}$ in the $y$ direction for a 12-chain of identical spheres with $d=5a$ for the optimal mode of planar flagellar swimming, labeled as in Fig. 3.

\subsection*{Fig. 5}
Plot of the maximum efficiency $E_{Tmax}$ of a $N$-chain of identical beads of radius $a$ and a head of radius $b$ with rest distances $d,...,d,b+d$ and $d=5a,\;b=10a$ for longitudinal squirming (large dots) and transverse planar flagellar swimming (small dots) for $N=3,...16$.

\subsection*{Fig. 6}
Plot of the displacements $\delta_{jx}$ in the $x$ direction for a 11-chain with $d=5a,\;b=10a$ for the optimal mode of longitudinal squirming, labeled as in Fig. 3.

\subsection*{Fig. 7}
Plot of the displacements $\delta_{jy}$ in the $y$ direction for a 11-chain with $d=5a,\;b=10a$ for the optimal mode of planar flagellar swimming, labeled as in Fig. 3.

\subsection*{Fig. 8}
Plot of the maximum efficiency $E_{Tmax}$ of a $3C$-chain of identical beads of radius $a$ for rest distance $d=5a$ for longitudinal squirming (drawn curve) and transverse planar flagellar swimming (dashed curve) as a function of the stiffness $\sigma$.

\subsection*{Fig. 9}
Plot of the product $(b^2/a^2)E_{Tmax}$ of a 11-chain with $d=5a$ and head of radius $b$ for longitudinal squirming (drawn curve) and transverse planar flagellar swimming (dashed curve) as a function of the ratio $b/a$.

\subsection*{Fig. 10}
Plot of the $x$ component of the displacement of the first (upper curve) and last (lower curve) sphere of a $4C$-chain with $d=5a,\;b=10a,\;\sigma=10$ as a function of time.

\subsection*{Fig. 11}
Plot of the $y$ component of the displacement of the first (strong oscillations) and last (weak oscillations) sphere of a $4C$-chain with $d=5a,\;b=10a,\;\sigma=10$ as a function of time.

\subsection*{Fig. 12}
As in Fig. 10 for a $11C$-chain.

\subsection*{Fig. 13}
As in Fig. 11 for a $11C$-chain.

\subsection*{Fig. 14}
Plot of the ratio $\overline{U}_{sw}/\overline{U}_{sw2}$ for a $4C$-chain (drawn curve) and a $11C$-chain (dashed curve) as functions of $\log_{10}(\varepsilon)$, as described at the end of Sec. VI.

\newpage
\setlength{\unitlength}{1cm}
\begin{figure}
 \includegraphics{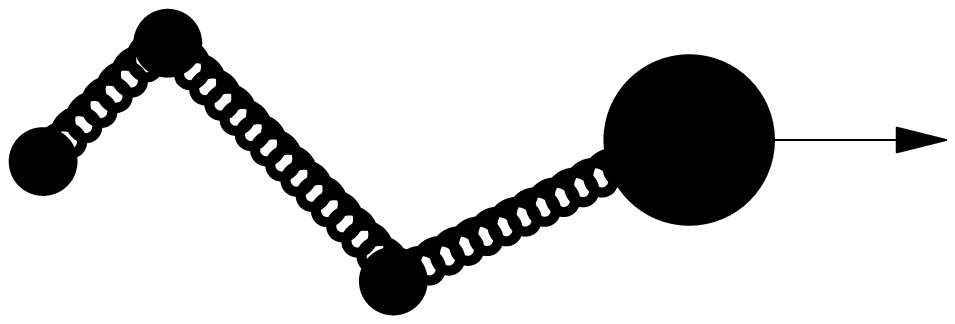}
   \put(-9.1,3.1){}
\put(-1.2,-.2){}
  \caption{}
\end{figure}
\newpage
\clearpage
\newpage
\setlength{\unitlength}{1cm}
\begin{figure}
 \includegraphics{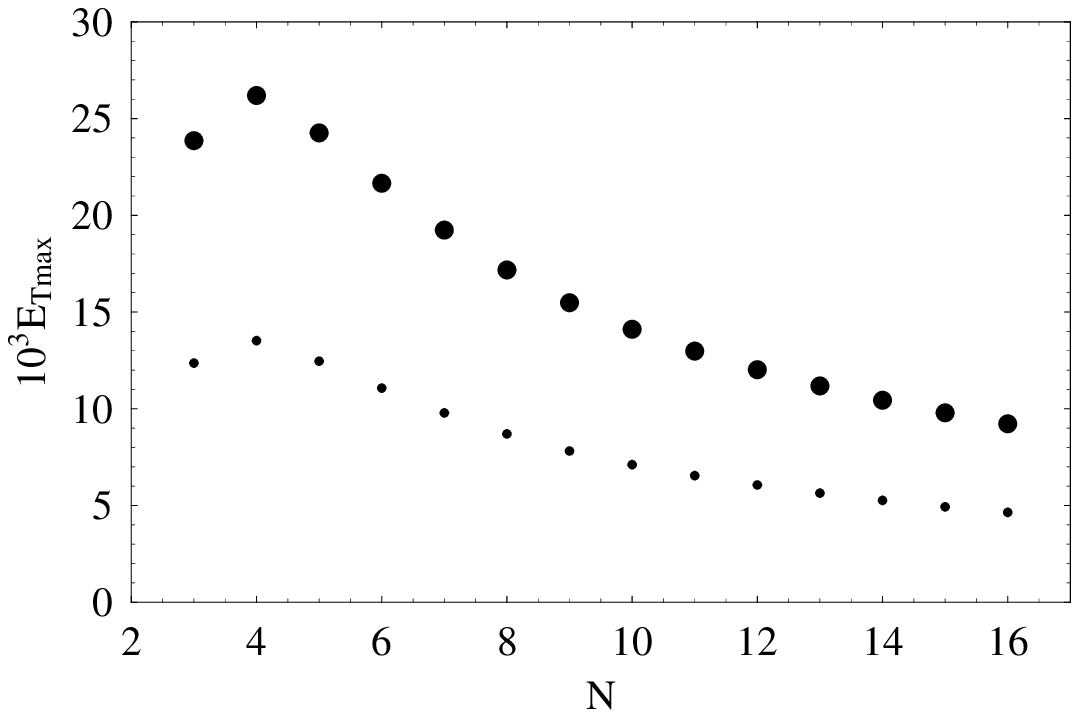}
   \put(-9.1,3.1){}
\put(-1.2,-.2){}
  \caption{}
\end{figure}
\newpage
\clearpage
\newpage
\setlength{\unitlength}{1cm}
\begin{figure}
 \includegraphics{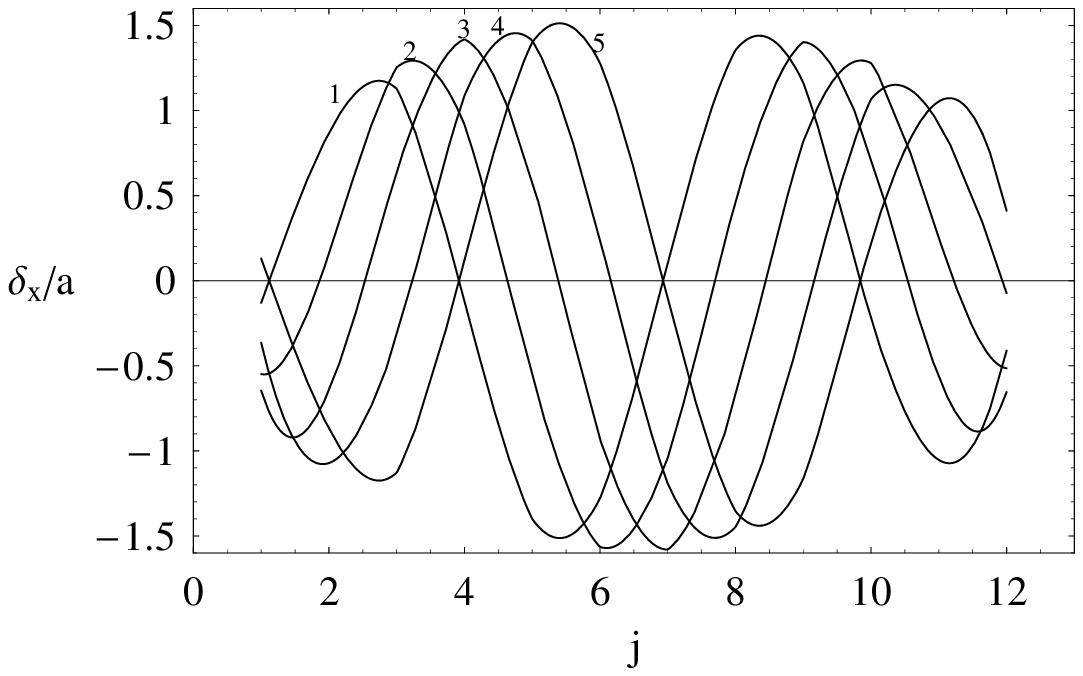}
   \put(-9.1,3.1){}
\put(-1.2,-.2){}
  \caption{}
\end{figure}
\newpage
\clearpage
\newpage
\setlength{\unitlength}{1cm}
\begin{figure}
 \includegraphics{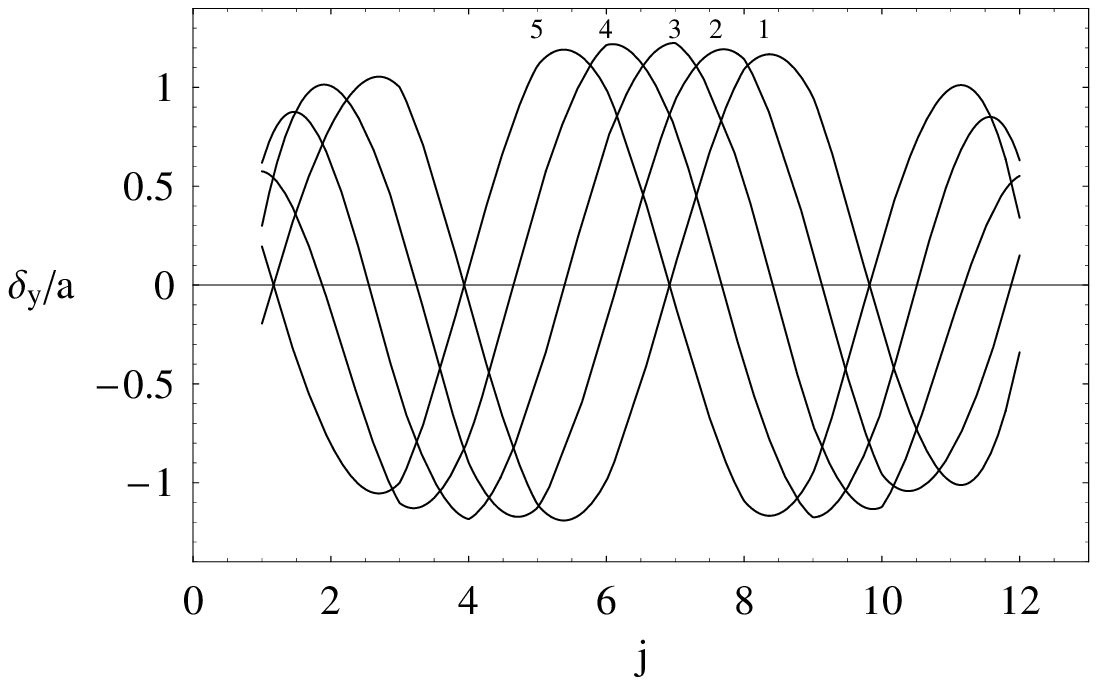}
   \put(-9.1,3.1){}
\put(-1.2,-.2){}
  \caption{}
\end{figure}
\newpage
\clearpage
\newpage
\setlength{\unitlength}{1cm}
\begin{figure}
 \includegraphics{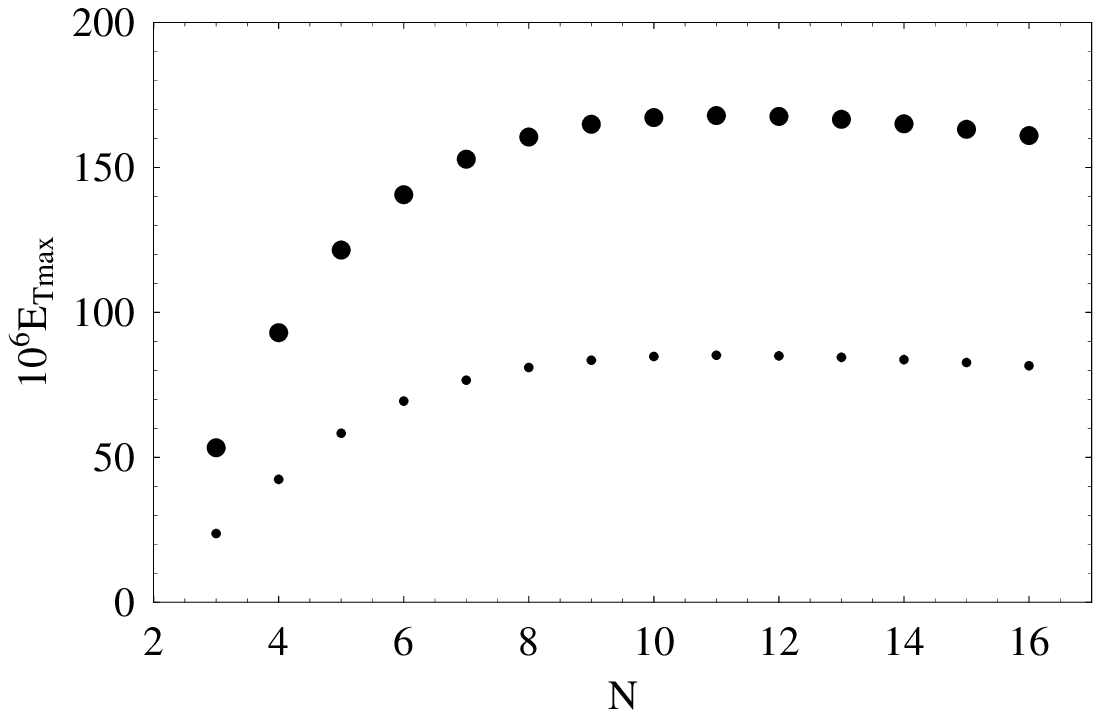}
   \put(-9.1,3.1){}
\put(-1.2,-.2){}
  \caption{}
\end{figure}
\newpage
\clearpage
\newpage
\setlength{\unitlength}{1cm}
\begin{figure}
 \includegraphics{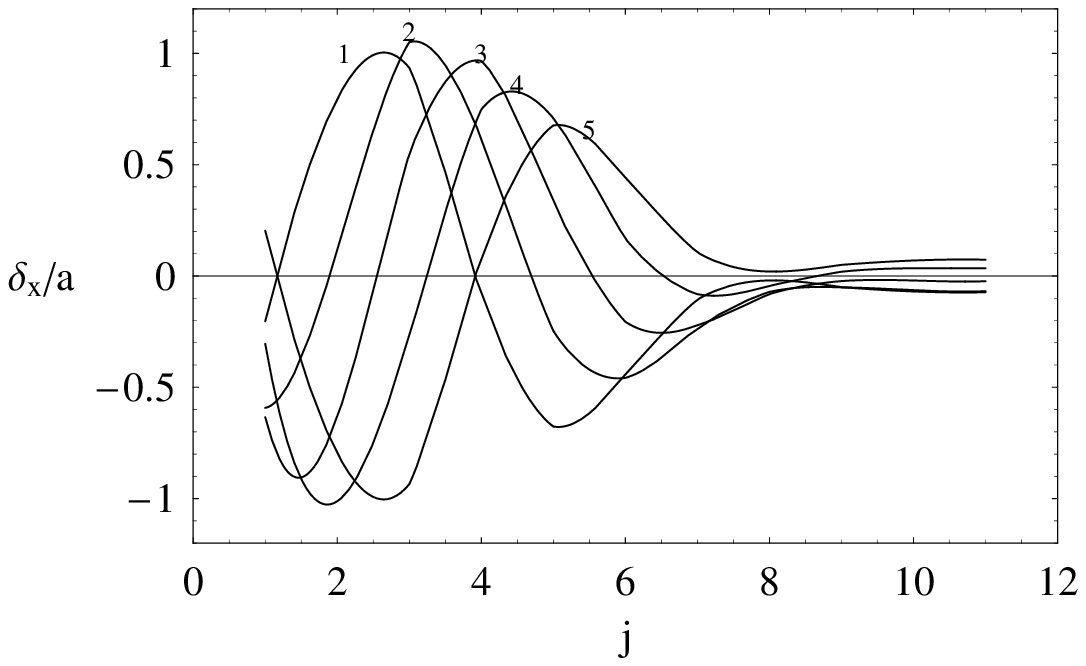}
   \put(-9.1,3.1){}
\put(-1.2,-.2){}
  \caption{}
\end{figure}
\newpage
\clearpage
\newpage
\setlength{\unitlength}{1cm}
\begin{figure}
 \includegraphics{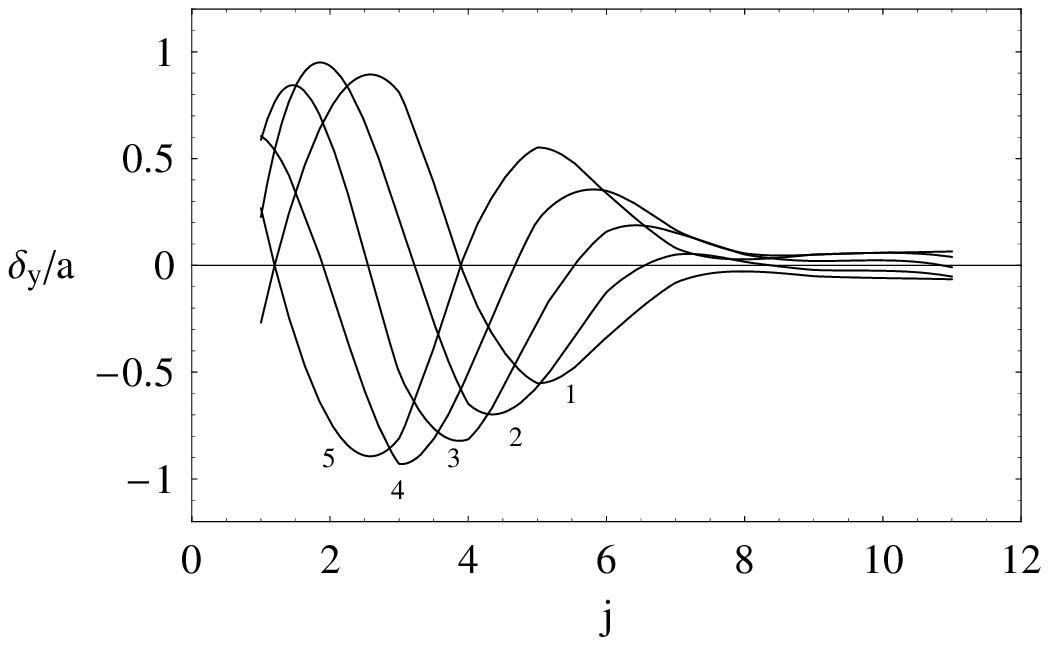}
   \put(-9.1,3.1){}
\put(-1.2,-.2){}
  \caption{}
\end{figure}
\newpage
\clearpage
\newpage
\setlength{\unitlength}{1cm}
\begin{figure}
 \includegraphics{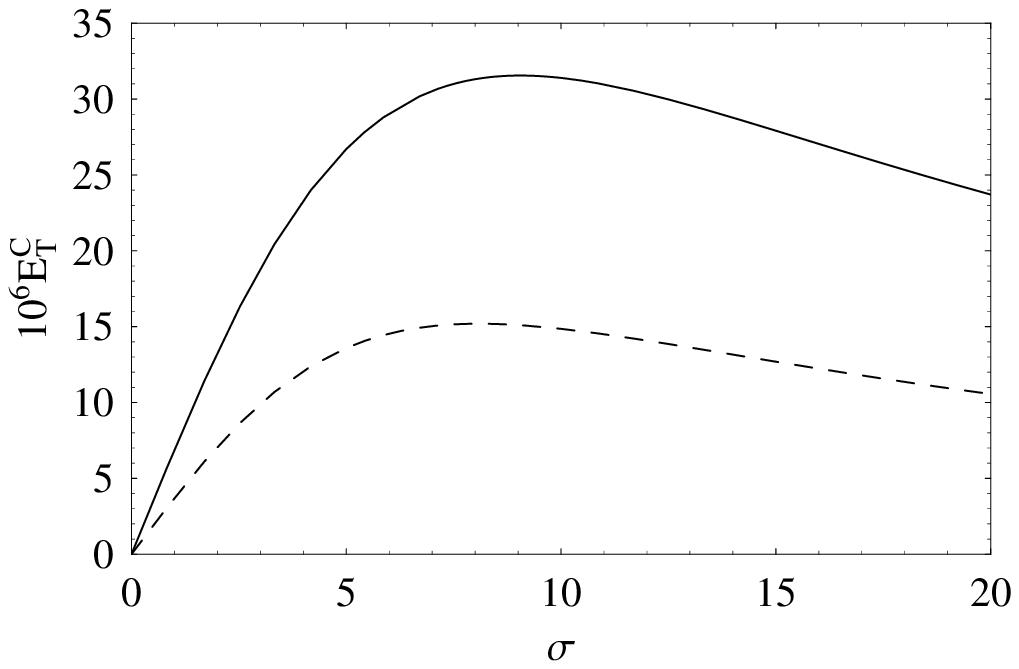}
   \put(-9.1,3.1){}
\put(-1.2,-.2){}
  \caption{}
\end{figure}
\newpage
\clearpage
\newpage
\setlength{\unitlength}{1cm}
\begin{figure}
 \includegraphics{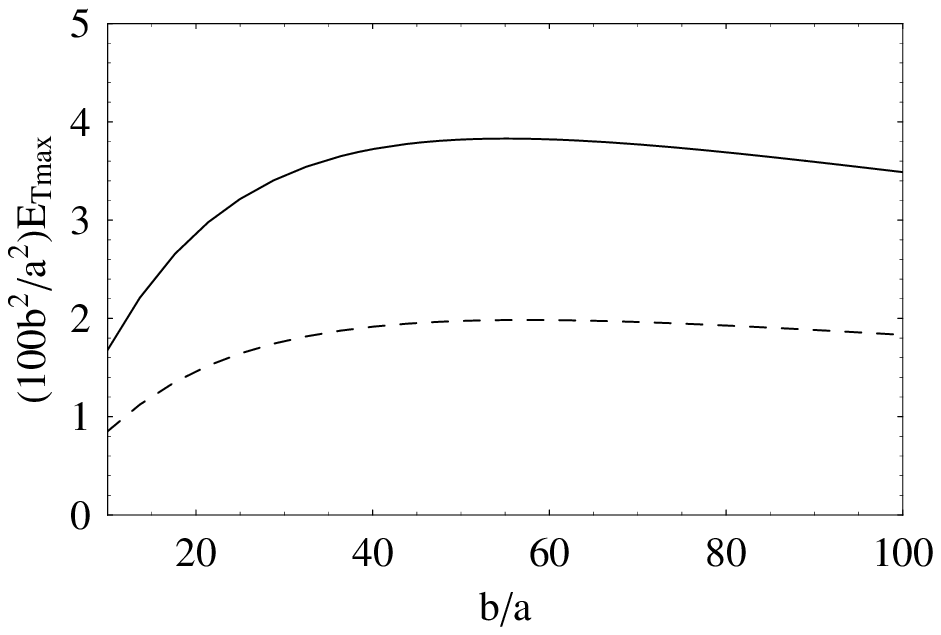}
   \put(-9.1,3.1){}
\put(-1.2,-.2){}
  \caption{}
\end{figure}
\newpage
\clearpage
\newpage
\setlength{\unitlength}{1cm}
\begin{figure}
 \includegraphics{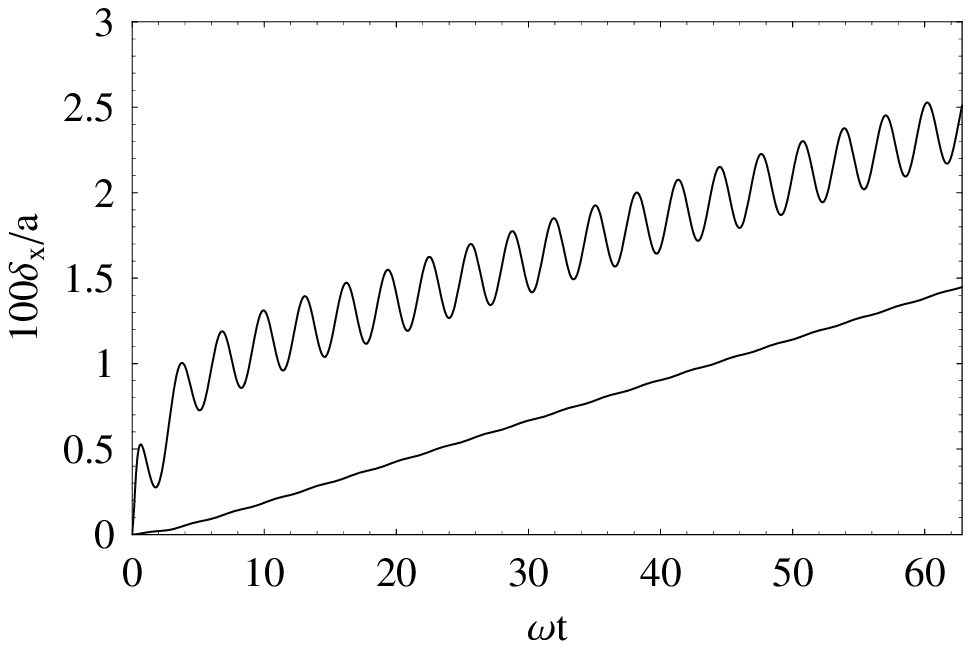}
   \put(-9.1,3.1){}
\put(-1.2,-.2){}
  \caption{}
\end{figure}
\newpage
\clearpage
\newpage
\setlength{\unitlength}{1cm}
\begin{figure}
 \includegraphics{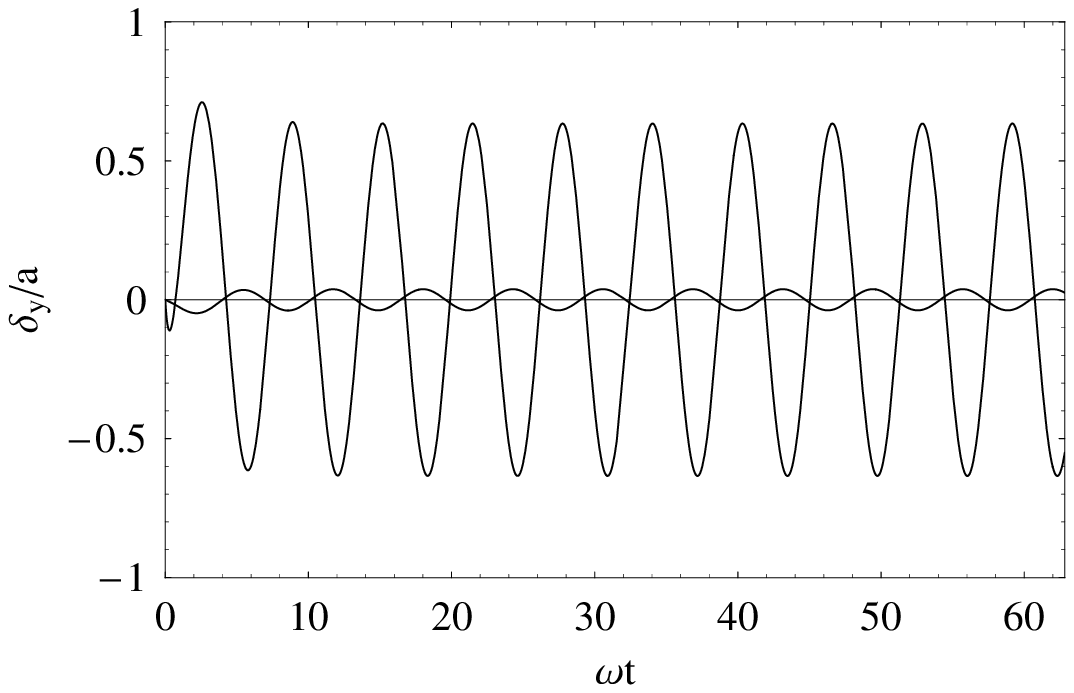}
   \put(-9.1,3.1){}
\put(-1.2,-.2){}
  \caption{}
\end{figure}
\newpage
\clearpage
\newpage
\setlength{\unitlength}{1cm}
\begin{figure}
 \includegraphics{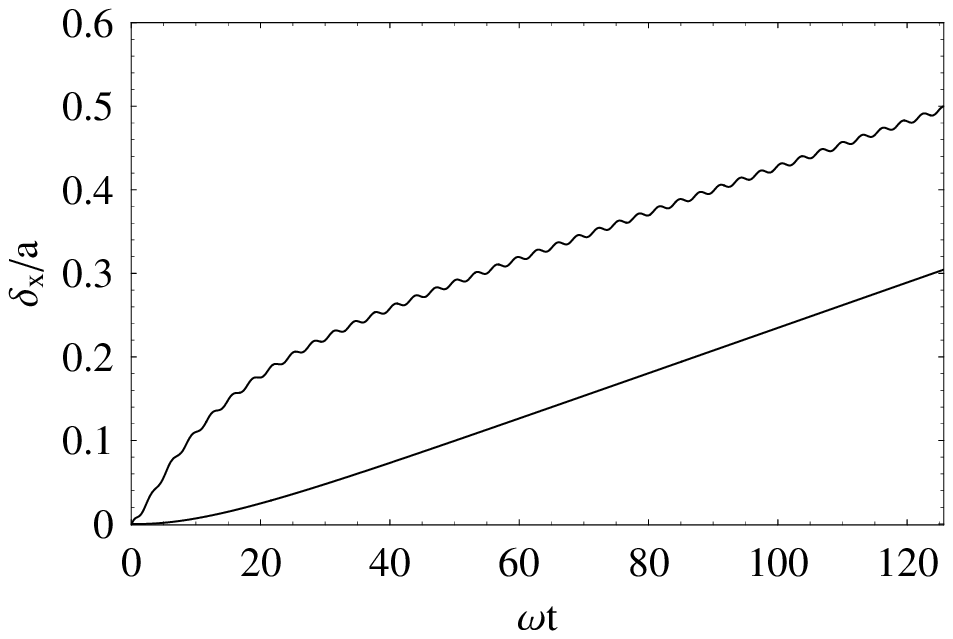}
   \put(-9.1,3.1){}
\put(-1.2,-.2){}
  \caption{}
\end{figure}
\newpage
\clearpage
\newpage
\setlength{\unitlength}{1cm}
\begin{figure}
 \includegraphics{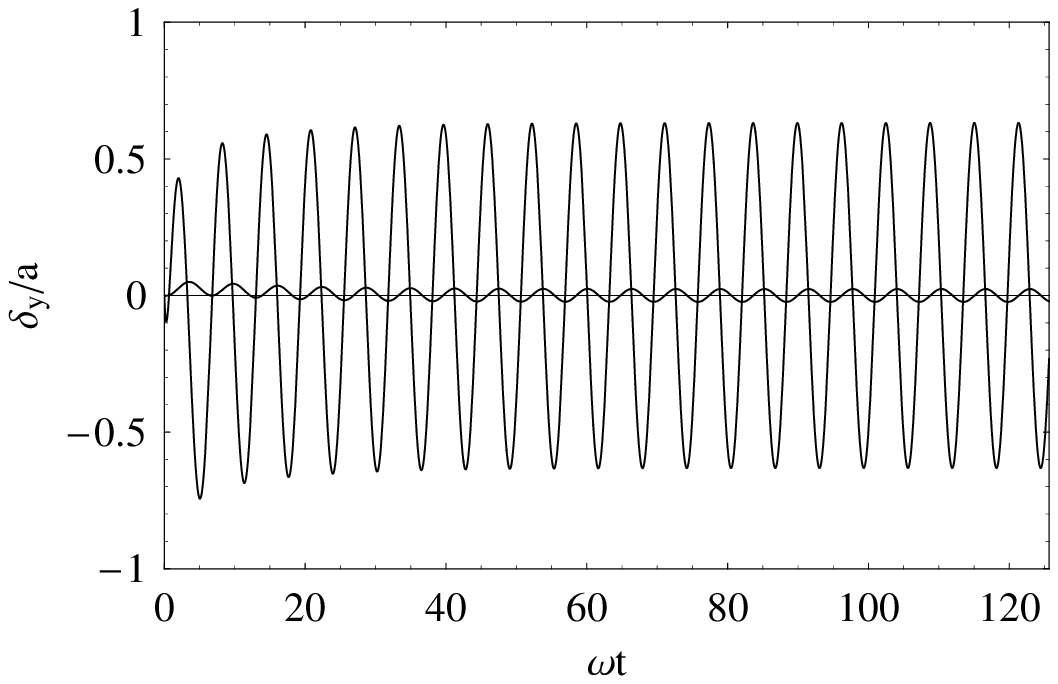}
   \put(-9.1,3.1){}
\put(-1.2,-.2){}
  \caption{}
\end{figure}
\newpage
\clearpage
\newpage
\setlength{\unitlength}{1cm}
\begin{figure}
 \includegraphics{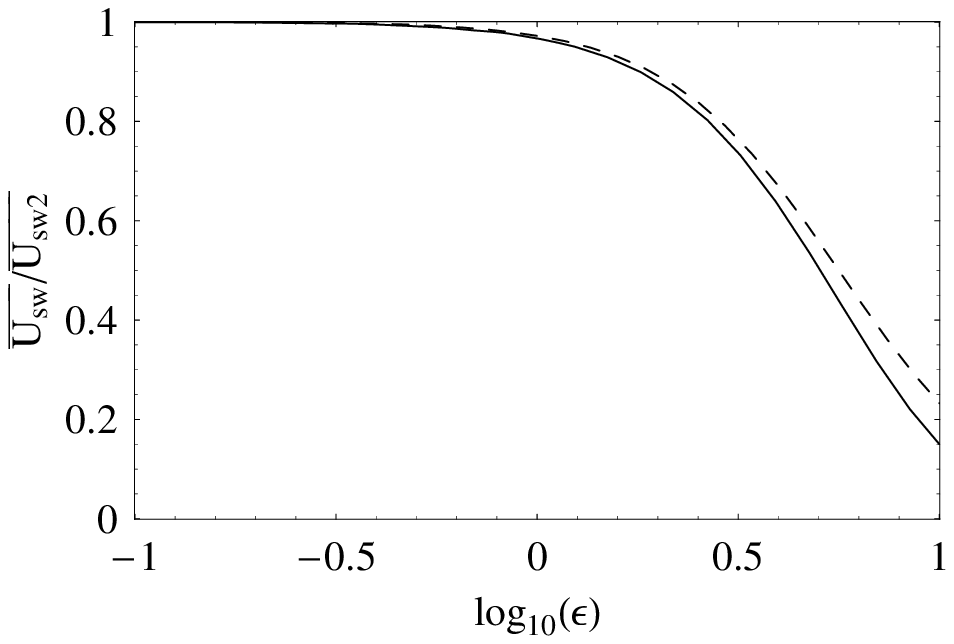}
   \put(-9.1,3.1){}
\put(-1.2,-.2){}
  \caption{}
\end{figure}
\newpage

\end{document}